\documentstyle[12pt,epsf,epsfig]{article}

\newlength{\dinwidth}
\newlength{\dinmargin}
\begin{document}

\def\bold#1{\setbox0=\hbox{$#1$}%
     \kern-.025em\copy0\kern-\wd0
     \kern.05em\copy0\kern-\wd0
     \kern-.025em\raise.0433em\box0 }
\def\slash#1{\setbox0=\hbox{$#1$}#1\hskip-\wd0\dimen0=5pt\advance
       \dimen0 by-\ht0\advance\dimen0 by\dp0\lower0.5\dimen0\hbox}
\newcommand{\dd}{\displaystyle}
\newcommand{\nn}{\nonumber}
\newcommand{\be}{\begin{equation}}
\newcommand{\ee}{\end{equation}}
\newcommand{\bea}{\begin{eqnarray}}
\newcommand{\eea}{\end{eqnarray}}	
\thispagestyle{empty}
\vspace*{1cm}
\rightline{BARI-TH/99-368}
\rightline{CERN-TH/99-403}
\rightline{UGVA-DPT 1999/11-1057}
%\rightline{November 1999}
\vspace*{2cm}
\begin{center}
  \begin{Large}
{\bf B Meson Transitions into Higher Mass Charmed Resonances}\\
  \end{Large}
\vspace{8mm}
  \begin{large}
P. Colangelo$^a$, F. De Fazio$^{b}$ 
\footnote{``Fondazione Angelo Della Riccia'' Fellow. 
Address  after December 1st, 1999: Centre for Particle Physics, 
Durham University, United Kingdom}
and G. Nardulli$^{a,c,d}$\\
  \end{large}
  \vspace{6mm}
{\it $^a$ Istituto Nazionale di Fisica Nucleare, Sezione di Bari, Italy}\\
{\it $^b$ D\'epartement de Physique Th\'eorique, Univ. de Gen\`eve, 
Switzerland}\\
{\it $^c$ Dipartimento di Fisica, Universit\'a di Bari, Italy}\\
{\it $^d$ Theory Division, CERN,  Gen\`eve, Switzerland}\\
\end{center}

\begin{quotation}
\vspace*{1.5cm}
\begin{center}
  \begin{bf}
  Abstract\\
  \end{bf}
\end{center}
\noindent
We use  QCD sum rules to  estimate the universal form factors
describing the semileptonic  
$B$ decays into excited charmed  resonances, such as the
$1^-$ and $2^-$ states $D_1^{*}$ and $D_2^{*}$ belonging to the
$\dd s_\ell^P=\frac{3}{2}^-$ heavy quark doublet, and
the $2^-$ and $3^-$ states $D_2^{*\prime}$ and $D_3$ belonging to the
$\dd s_\ell^P=\frac{5}{2}^-$ doublet. 
\vspace*{0.5cm}
\end{quotation}
\newpage
\baselineskip=18pt
\setcounter{page}{1}

\section{Introduction}    

In the heavy quark ($Q=c,b$) infinite mass limit ($m_Q \to \infty$) Quantum
Chromodynamics exhibits symmetries that are not present in the finite 
mass theory:   heavy quark spin and flavour symmetries
\cite{hqet}, as well as the velocity superselection rule \cite{georgi}.
 These approximate symmetries allow  to organize the spectrum of physical
states comprising one light antiquark and one heavy quark in multiplets of
definite parity $P$ and total angular momentum  $s_\ell$
of the light degrees of freedom. 

The lowest lying multiplet consists in the meson doublet with
$s^P_\ell \, = \; {\frac{1}{2}}^-$, corresponding to the vector $1^-$ and 
the pseudoscalar $0^-$ state. The doublet can be described by a  
$4 \times 4$ Dirac matrix 
\begin{equation}
H = \frac{(1+{\rlap{v}/})}{2}[P_{\mu}^*\gamma^\mu-P\gamma_5] 
\label{h}
\end{equation}
where $v$ is the heavy meson velocity, 
$P^{*\mu}_a$ and $P_a$ are annihilation operators 
of the $1^-$ and $0^-$ $Q{\bar q}_a$ mesons 
($a=1,2,3$ for $u,d$ and $s$); for charm, they are $D^*$ and $D$,
respectively.
\footnote{The operators in (\ref{h}) 
 have dimension $\frac{3}{2}$ since they contain a factor 
${\sqrt {m_P}}$ in their definition.}

The nearest mass multiplets are the $s^P_\ell \, = \; {\frac{1}{2}}^+$ 
doublet, comprising  the positive parity $1^+$ and $0^+$ states,
and the $s^P_\ell \, = \; {\frac{3}{2}}^+$ 
doublet which includes  the positive parity $1^+$ and $2^+$
states. In the charm sector three of such states have been identified:
the state $D_2(2460)$ is the narrow   $2^+$ meson  with 
$s^P_\ell \, = \; {\frac{3}{2}}^+$; moreover, there are two $1^+$ mesons with
masses  $m_{D_1^{0}}=(2422.2\pm1.8)$ MeV \cite{pdg}
and $m_{D_1^{*0}}=(2461^{+41}_{-34}\pm10\pm32)$ MeV \cite{cleod}; 
they can be identified with members of the multiplets
predicted by the Heavy Quark Effective Theory  \cite{falkluke}, 
including some mixing between them. Evidence for such states has also been
collected in the beauty sector \cite{ciulli}.
From the theoretical viewpoint these states have been the subject
of intense scrutiny:
the role of the ${\frac{1}{2}}^+$ doublet  $(0^+  , \; 1^+)$ in some 
applications of chiral perturbation theory has been considered in 
\cite{falkluke} and in \cite{falk};
their properties have been studied both by
QCD sum rules \cite{paver,gatto,dai} and quark models \cite{qmod}.

In this letter we  investigate some properties of the
next heavy meson multiplets, 
the $s^P_\ell \, = \; {\frac{3}{2}}^-$ doublet including two mesons with
$J^P=1^-~$ and $2^-$, and the  $s^P_\ell \, = \; {\frac{5}{2}}^-$ 
doublet which comprises the  states with $J^P=2^-~$ and $3^-$. 
We estimate the universal form 
factors describing, in the infinite heavy quark mass limit, 
 the semileptonic $B$ decays into such multiplets, and 
consider the contribution of these processes to the inclusive semileptonic 
$B$ decay width 
\footnote{A review on the problems related to inclusive and 
exclusive semileptonic B decays can be found in ref.\cite{babar}.}. 

We  follow the QCD sum rule approach \cite{svz},
which has been applied to similar problems in the past \cite{paver,dai,neub} 
\footnote{For a review see \cite{neubertrev}.}. 
However, as  discussed in the following, in the application of
the method to high-spin 
states several difficulties appear 
in identifying the range of parameters needed in the sum 
rule analyses, due to the peculiar features of the considered states 
and of their interpolating currents. 
In order to overcome such difficulties, we make use of information 
coming from other theoretical approaches, namely constituent quark models
predicting the heavy meson spectrum. The final result, although
affected by a sizeable theoretical uncertainty, nevertheless is 
useful for assessing the  role of high-spin meson doublets in
constituting part of the charm inclusive semileptonic B decay width.

\section{Effective meson operators and quark currents}    
The effective operators describing the  $s^P_\ell \, = \; {\frac{3}{2}}^-$ and
$s^P_\ell \, = \; {\frac{5}{2}}^-$ meson doublets are  given
respectively by \cite{falk}:
\bea
H^\mu&=&\frac{1+{\rlap{v}/}}{2}
\left[D_2^{*\mu\nu}\gamma_5\gamma_\nu-\sqrt{\frac{3}{2}}D_1^{*\nu}
(g^\mu_\nu-\frac{\gamma_\nu}{3}(\gamma^\mu+v^\mu)\right]
\label{states1}
\\
H^{\mu\nu}&=&\frac{1+{\rlap{v}/}}{2}
\left[D_3^{\mu\nu\sigma}\gamma_\sigma-\sqrt{\frac{5}{3}}\gamma_5 
D_2^{*\prime\alpha\beta}
\left(g^\mu_\alpha
g^\nu_\beta-
\frac{\gamma_\alpha}{5}g^\nu_\beta (\gamma^\mu-v^\mu)-
\frac{\gamma_\beta}{5}g^\mu_\alpha (\gamma^\nu-v^\nu)\right)\right] 
\;,\;\;
\label{states2}
\eea
where $D^*_i$ represent annihilation operators of the mesons with appropriate 
quantum numbers.
In order to implement the QCD sum rule programme, we need quark
currents with non-vanishing projection on 
these states. They have been investigated in ref.\cite{dai} and are
given by the following expressions:
\bea
&s_\ell^P=(\frac{3}{2})^-;~J^P=1^-&:~~~~~~~
J^\alpha=-\bar{h}_v\sqrt{\frac{3}{4}}
\left[D^\alpha_t~-~\frac{\gamma^\alpha_t}{3}\, {\rlap{D}/}_{\, t}\right]q
\label{ja}\\
&s_\ell^P=(\frac{3}{2})^-;~J^P=2^-&:~~~~~~
J^{\alpha\beta}=T^{\alpha\beta,\, \mu\nu}\bar{h}_v~
\left[\frac{1}{\sqrt{2}}~\gamma_5
\gamma_{t\, \mu}
D_{t\, \nu}\right] q\\
&s_\ell^P=(\frac{5}{2})^-;~J^P=2^-&:~~~~~~
\tilde{J}^{\alpha\beta}=-\sqrt{\frac{5}{6}}\,
T^{\alpha\beta ,\, \mu\nu}\bar{h}_v~\gamma_5\,
\left[D_{t\, \mu} 
D_{t\, \nu}~-~\frac{2}{5}\, 
D_{t\, \mu}     \gamma_{t\, \nu} \,     {\rlap{D}/}_{\, t}   \right]q
\label{jabt}\\
&s_\ell^P=(\frac{5}{2})^-;~J^P=3^-&:~~~~~
{J}^{\alpha\beta\lambda}=T^{\alpha\beta\lambda
,\,  \mu\nu\sigma}\bar{h}_v\left[
\frac{i}{\sqrt{2}}~\gamma_{t\, \mu}D_{t\, \nu}D_{t\, \sigma}\right]q~,
\label{jabc}\eea
where $D^\mu$ is the covariant derivative:
$D^\mu~=~\partial^\mu~-~i\, g\, A^\mu~$, and
$G^\mu_t$ represents the transverse component of the four-vector
$G^\mu$ with respect to the heavy quark velocity $v$:
$G^\mu_t~=G^\mu~-~(G \cdot v) v^\mu~$.
The tensors $T^{\alpha\beta,\, \mu\nu}$ 
and $T^{\alpha\beta\lambda,\, \mu\nu\sigma}$ are needed to symmetrize indices
and are given by
\begin{eqnarray}
T^{\alpha\beta ,\, \mu\nu}&=&	
\frac{1}{2}\left(g^{\alpha\mu}g^{\beta\nu}+
g^{\alpha\nu}g^{\beta\mu}\right)-\frac{1}{3}\,
g_t^{\alpha\beta}g_t^{\mu\nu}\\
T^{\alpha\beta\lambda ,\, \mu\nu\sigma}&=&\frac{1}{3}\left(
g^{\alpha\mu}g^{\beta\nu}g^{\lambda\sigma}+
g^{\alpha\nu}g^{\beta\mu}g^{\lambda\sigma}+
g^{\alpha\sigma}g^{\beta\nu}g^{\lambda\mu}\right)\nn
\\
&-&\frac{1}{3}\,
\left(
g_t^{\alpha\beta}g_t^{\mu\nu}g_t^{\lambda\sigma}+
g_t^{\alpha\lambda}g_t^{\mu\nu}g_t^{\beta\sigma}+
g_t^{\beta\lambda}g_t^{\mu\nu}g_t^{\alpha\sigma}\right)~,
\end{eqnarray}
with
$g_t^{\alpha\beta}~=~g^{\alpha\beta}~-~v^\alpha v^\beta~$.

As discussed in \cite{dai}, 
in the $m_Q\to\infty$ limit the currents in eqs.(\ref{ja})-(\ref{jabc})
 have non-vanishing
projection only to the corresponding states of the HQET, without
mixing with states of the same quantum number but different $s_\ell$
content. Therefore, we can define a set of one-particle-current couplings as
follows:
\bea
&s_\ell^P=(\frac{3}{2})^-;~J^P=1^-&:~~~~~~~~~~~~~
<D^{*}_1(v,~\epsilon)~|J^\alpha|~0>~=~
f_1~\sqrt{m_{D^*_1}}~\epsilon^{*\alpha} \\
&s_\ell^P=(\frac{3}{2})^-;~J^P=2^-&:~~~~~~~~~~~~
<D^{*}_2(v,~\epsilon)~|J^{\alpha\beta}|~0>~=~
f_2~\sqrt{m_{D^*_2}}~\epsilon^{*\alpha\beta} \\
&s_\ell^P=(\frac{5}{2})^-;~J^P=2^-&:~~~~~~~~~~~~
<D^{*\prime}_2(v,~\epsilon)~|\tilde{J}^{\alpha\beta}|~0>~=~
\tilde{f}_2~\sqrt{m_{D^{*\prime}_2}}~\epsilon^{*\alpha\beta}\\
&s_\ell^P=(\frac{5}{2})^-;~J^P=3^-&:~~~~~~~~~~~
<D^{*}_3(v,~\epsilon)~|\tilde{J}^{\alpha\beta\lambda}|~0>~=~ 
f_3~\sqrt{m_{D^*_3}}~\epsilon^{*\alpha\beta\lambda}~,
\eea where $\epsilon$ are the meson polarization tensors. 
The couplings $f_i$ are low-energy parameters, determined by the dynamics of 
the light degrees of freedom.
Since the two pairs $(f_1,~f_2)$ and $(\tilde{f}_2,~f_3)$ are related by the 
spin symmetry, in
the sequel we only consider $f_1$ and $\tilde{f}_2$.

\section{Two-point function sum rules}
To evaluate the  parameters  $f_1$ and $\tilde{f}_2$ let us consider
the two-point correlators
\bea
i~\int d^4 x~e^{-i\omega v \cdot x}
<0~|T(J^{\dag\alpha}(x)J^\beta(0)|~0>&=&\Pi_1(\omega)
g_t^{\alpha\beta} \label{corr1}\\
i~\int d^4 x~e^{-i\omega v \cdot x}
<0~|T(\tilde{J}^{\dag\alpha\beta}(x)\tilde{J}^{\mu\nu}(0)|~0>&=&
\Pi_2(\omega)\left(g_t^{\alpha\mu}g_t^{\beta\nu}\, +\, 
g_t^{\alpha\nu}g_t^{\beta\mu}\, -\, \frac{2}{3}\, 
g_t^{\alpha\beta}g_t^{\mu\nu}
\right) \nonumber \label{corr2}\\
\eea
given in terms of 
$\Pi_1$ and $\Pi_2$, scalar functions of the variable $\omega$.

As extensively discussed in the literature, 
the QCD sum rule method amounts to evaluate the correlators in two equivalent
ways. On one side the  Operator Product Expansion
(OPE) is applied for negative values of $\omega$; the expansion produces
an asymptotic series,
whose leading term is the perturbative contribution (computed in HQET), 
followed by subleading
terms parameterized by non perturbative quantities, such as the quark
condensate: $<\bar q q	>$, the gluon condensate:  
$<\alpha_s G_{\mu\nu}G^{\mu\nu}>$, the mixed quark-gluon condensate, etc. 
On the other side, one
evaluates the correlators by writing down dispersion relations (DR) for the
scalar functions $\Pi_1(\omega)$ and $\Pi_2(\omega)$; 
they get contributions by the hadronic
states, in particular by the low-lying 
resonances with  appropriate quantum numbers. To get rid of radial
excitations and multiparticle states, one performs a
Borel transform on both sides of the sum rule,  which 
enhances the low mass contribution of the
spectrum; moreover, assuming quark-hadron duality, one identifies,
from some effective continuum 
threshold $\omega_c$, the hadronic side of the sum rules 
with the perturbative result obtained by the OPE. In the final sum
rule, only the contributions from the physical to the continuum
threshold appear: the low mass resonance on one side, the OPE truncated 
at $\omega_c$ on the other.

Applying the method to the correlators (\ref{corr1}) and (\ref{corr2})
we get two borelized sum rules for the  parameters $f_1$ and $\tilde f_2$:
\bea
f^2_1 e^{-\Delta_1/E} &=&  \frac{2}{\pi^2}\, \int_0^{\omega_{1\, c}} 
d \sigma~\sigma^4 e^{-\sigma/E}~+~\frac{<\bar q\sigma G q>}{16}
\label{2p1}
\\
\tilde{f}^2_2 e^{-\Delta_2/E'} &=&
\frac{16}{5 \pi^2 }\int_0^{\omega_{2\,c}}
d \sigma~\sigma^6 e^{-\sigma/E'}~.
\label{2p2}
\eea
Here the parameters $\Delta_1$ and $\Delta_2$ are defined
by the formulae:
 $\Delta_1=m_{D^*_1}-m_c $ and $\Delta_2=m_{D^{*\prime}_2}-m_c$,
$m_c$ being the charm quark mass; therefore, the parameters 
$\Delta_1$ and $\Delta_2$ represent
the binding energy of the states $D^*_1$ and   $D^{*\prime}_2$, which is finite
in the infinite heavy quark mass limit. On the other hand,
$\omega_{1\, c}$ and 
$\omega_{2\, c}$ represent the effective thresholds
separating the low-lying resonances from the continuum;
$E$ and $E'$ are parameters introduced by the Borel procedure. 
Relations for   the mass parameters $\Delta_1$ and $\Delta_2$ can be
obtained by taking derivatives of the  sum rules (\ref{2p1}) and (\ref{2p2}):
\bea
\Delta_1 &=&\frac{\dd \frac{1}{\pi^2}\int_0^{\omega_{1\,c}}
d \sigma~\sigma^5 e^{-\sigma/E}}{\dd  \frac{1}{\pi^2} \int_0^{\omega_{1\,c}}
d \sigma~\sigma^4 e^{-\sigma/E} ~+~\frac{<\bar q\sigma G q>}{32} }
\label{d1}\\
&&\nn\\
\Delta_2 &=&\frac{\dd \int_0^{\omega_{2\,c}}
d \sigma~\sigma^7 e^{-\sigma/E' }}{\dd \int_0^{\omega_{2\,c}}
d \sigma~\sigma^6 e^{-\sigma/E'} } ~.\label{d2}
\eea

There is an important point deserving a  discussion, and it concerns 
the high dimensionality of the interpolating currents $J^\alpha$ 
and $\tilde J^{\alpha \beta}$, which 
has two consequences on the structure of the sum rules 
(\ref{2p1})-(\ref{2p2}) and (\ref{d1})-(\ref{d2}). 
First, the spectral functions in 
eqs.(\ref{2p1})-(\ref{2p2}) and (\ref{d1})-(\ref{d2}) have large powers, 
and therefore the perturbative contributions in the sum rules 
are very sensitive to the continuum thresholds $\omega_{1\,c}$ and 
$\omega_{2\,c}$. The second effect consists in the absence of the contributions
from low-dimensional condensates, which implies
(neglecting high-dimensional condensates) complete 
duality between the perturbative and the hadronic contributions to the
sum rules. Such two effects cannot be avoided in our analysis, 
and are typical
of the sum rule approach to high spin states \cite{reinders}. In our case they
have the main consequence of not allowing  to 
determine simultaneously
 the couplings $f_i$ and the mass parameters $\Delta_i$, due to
the critical dependence on the continuum thresholds. Therefore, we adopt
the strategy of getting the values of the mass parameters from other
determinations, and then to fix the thresholds from eqs.(\ref{d1})-(\ref{d2})
and computing $f_i$ from (\ref{2p1})-(\ref{2p2}). 
Admittedly, this is
a hybrid procedure, which nevertheless allows us to estimate
both the current-particle matrix elements and the universal semileptonic
form factors, as
discussed in the next Section. 

While experimental information on the 
$s_\ell^P=\frac{3}{2}^-$ and
$s_\ell^P=\frac{5}{2}^-$
doublets are not available so far,
there are studies concerning such states 
based on constituent quark models \cite{godfrey}.
They suggest that the mass of the $3^-$ $(c \bar u)$ state $D_3$ is 
$m_{D_3}=2.83$ GeV or $m_{D_3}=2.76$ GeV,
whereas  the mass of the corresponding 
$(b \bar u)$ state is $m_{B_3}=6.11$ GeV. 
Assuming a spin splitting of 
$\simeq 40$ MeV in the charm sector, as suggested by the same models, 
we can
give to the mass of the ${5 \over 2}^-$ state the value of $2.78$ GeV, e.g.
nearly $0.8$ GeV above the  $0^-$ doublet 
(the same value comes from the analysis of the beauty meson spectrum).
This implies for  the parameter $\Delta_2$ 
a value in the range $\Delta_2\simeq[1.2-1.4]$ GeV, 
considering the determination of the analogous binding
energy of  $B$ and $D$ mesons \cite{neubertrev}. As for $\Delta_1$,
we fix it to  $\Delta_1\simeq[1.2-1.4]$ GeV, according to similar 
considerations.
  
Let us consider $\Delta_1$ and $\Delta_2$ related to the
thresholds $\omega_i$ and to the Borel parameters $E_i$ by
eqs.(\ref{d1})-(\ref{d2}). There is a range of
Borel parameters and thresholds where the chosen binding energies can be
obtained. In particular, while the dependence of $\Delta_i$ on the Borel 
parameters is quite mild, so that the range $E_i=[1-1.5]$ GeV can be chosen, 
the dependence on the thresholds, as expected, is critical: 
one has to choose $\omega_i$ in a quite narrow range $[1.6-1.8]$ GeV
to obtain $\Delta_i$. However, this choice is not
unappropriate, since it suggests that the
continuum  threshold is above the mass of the corresponding
resonance by nearly the  mass of one pion.

After having fixed $\Delta_i$ and  the ranges of $E_i$ and of $\omega_i$, from 
eqs.(\ref{2p1})-(\ref{2p2}) we can obtain the values of the couplings $f_i$:
$f_1=[0.6-0.8]$ GeV$^{5\over2}$ and 
$\tilde f_2=[1.2-1.6]$ GeV$^{7\over2}$.
Notice that, at odds, e.g., with the leptonic constants related to the 
matrix elements of the quark axial currents on the $0^-$ state, the 
couplings $f_i$ do not have an immediate physical meaning, as they
represent the projections of the interpolating currents on the 
orbitally excited meson states. 
Nevertheless, they 
play an important role in the determination of the form factors, as we  
discuss in the next Section.

\section{Universal form factors from three-point  sum rules}
There are two universal form factors describing the semileptonic $B$ decays
into the excited negative parity charmed resonances with 
$s_\ell={3\over 2}^-$ and $s_\ell={5\over 2}^-$. The first one,
$\tau_1$, governs the decays
\bea
&&B\rightarrow D^*_1\ell\nu_\ell\label{eq:4.1} \\
&&B\rightarrow D^*_2\ell\nu_\ell\label{eq:4.2}
\eea 
in the heavy quark limit. The second
one, $\tau_2$,  describes in the same limit the decays
 \bea
&& B \rightarrow D^{*\prime}_2 \ell \nu_{\ell}\label{eq:4.3} \\
&& B \rightarrow D_3 \ell \nu_\ell~.\label{eq:4.4}
\eea
It is straightforward to write down the semileptonic matrix elements
for the transitions (\ref{eq:4.1})-(\ref{eq:4.4}),
by applying, e.g., the trace formalism \cite{neubertrev}. One obtains:
\bea
< D^*_1 (v^\prime,\epsilon)|(V-A)^\mu|B(v)>&=&\sqrt{m_B m_{D^*_1   }  }~
\tau_1(y)~\sqrt{\frac{3}{2}}~\Big[
(\epsilon^*v)\left(v^\mu-v^{\prime\mu}~+~\frac{1-y}{3}\,
v^{\prime\mu}\right)\nn\\
&-&\frac{1-y^2}{3}\epsilon^{*\mu}+i
\frac{1-y}{3}\epsilon^{\mu\lambda\rho\sigma}
\epsilon^*_{\lambda}v^\prime_\rho v_\sigma 
\Big] ~,\label{eq:4.5}
\\ 
< D^*_2 (v^\prime,\epsilon)|(V-A)^\mu|B(v)>&=&\sqrt{m_B m_{D^*_2   }  }
~\tau_1(y)~\epsilon^*_{\lambda\nu}\, v^\lambda \Big[
g^{\mu\nu}(y-1)-v^\nu v^{\prime\mu}\nn\\
&+&~i~
\epsilon^{\alpha\beta\nu\mu}v^\prime_\alpha v_\beta
\Big]\label{eq:4.6}
\eea for the decays (\ref{eq:4.1}) and (\ref{eq:4.2}), while for the decays
(\ref{eq:4.3}) and (\ref{eq:4.4}) the relevant matrix elements can be 
written as:
\bea 
< D^{*\prime}_2
(v^\prime,\epsilon)|(V-A)^\mu|B(v)>&=&\sqrt{\frac{5}{3}}\, \sqrt{m_B
m_{D^{*\prime}_2 } } ~\tau_2(y)~\epsilon^*_{\alpha\beta}\, v^\alpha \Big[
\frac{2(1-y^2)}{5}\, g^{\mu\beta}-v^\beta v^{\mu}\nn\\
&+&\, \frac{2y-3}{5}\,v^\beta v^{\prime\mu}~+ ~i~\frac{2(1+y)}{5}
\epsilon^{\mu\lambda\beta\rho} v_\lambda  v^\prime_\rho 
\Big]~,\label{eq:4.7}\\
< D_3
(v^\prime,\epsilon)|(V-A)^\mu|B(v)>&=&\sqrt{m_B
m_{D_3 } } ~\tau_2(y)~\epsilon^*_{\alpha\beta\lambda}\, v^\alpha
v^\beta\Big[g^{\mu\lambda}(1+y)-v^\lambda v^{\prime\, \mu}\nn\\
&+&i \epsilon^{\mu\lambda\rho\tau} v_\rho v^\prime_\tau\Big]
~.\label{eq:4.8}
\eea In these equations the weak current is 
$(V-A)^\mu=\bar c\gamma^\mu(1-\gamma_5)b$, $y=v \cdot v^\prime$ and
$\tau_1(y)$, $\tau_2(y)$ are the universal form factors.

At the zero-recoil point $v =v^\prime$ the matrix elements in 
(\ref{eq:4.5})-(\ref{eq:4.8}) vanish, as expected by the heavy quark symmetry. 
As a matter of fact, for $B$ decays into spin 2 and spin 3 states, at least one
index of the final meson polarization tensor is contracted by the 
$B$ four-velocity $v$, and therefore the product vanishes for $v =v^\prime$. 
The spin symmetry
requirement being verified in the matrix elements, the Isgur-Wise form factors 
$\tau_1$ and $\tau_2$ are not required to vanish at 
$v \cdot v^\prime = 1$.

One can attempt an estimate of the form factors $\tau_{1,2}$ by three-point 
function sum rules, considering
the  correlators (relevant for  the matrix elements
(\ref{eq:4.5}) and (\ref{eq:4.7})):
\bea
i^2~\int d^4 x~d^4 z &&e^{-i(\omega
v\cdot x-\omega^\prime 
v^\prime \cdot z)} <0~|T(J^{\dag\alpha}(z)(V-A)^\mu(0) J_5(x)|~0>~=~
\nn\\&=&i\epsilon^{\mu\alpha\beta\lambda}v_\beta
v^\prime_\lambda\Omega_1(\omega,~\omega^\prime)+
\Xi_1(\omega,~\omega^\prime)w^\alpha v^\mu~+~...
\\
i^2~\int d^4 x~d^4 z &&e^{-i(\omega
v \cdot x-\omega^\prime 
v^\prime \cdot z)}<0~|T(\tilde{J}^{\dag\alpha\beta}(z)(V-A)^\mu(0)
J_5(x)|~0>~=\nn\\
&=&i\,
\epsilon^{\mu\sigma\tau\rho}v_\tau v^\prime_\rho (w^\alpha
g^\beta_\sigma \, +\,  w^\beta g^\alpha_\sigma)\Omega_2(\omega,~\omega^\prime)
\, +\, w^\alpha w^\beta v^\mu\Xi_2(\omega,~\omega^\prime)~+~...
\eea where $w^\alpha=v^\alpha-y v^{\prime\alpha}$, $J_5=\bar q
i\gamma_5 b$; the dots represent other Lorentz structures 
which are not relevant for the subsequent analysis, since we only consider
$\Omega_1$ and $\Omega_2$.

Since the scalar functions $\Omega_j$
depend on two variables, one has to perform  double DRs and
 double  Borel transforms, which introduces, for each sum rule, two 
Borel parameters $E$ and $E^\prime$.
The resulting equations read:
\bea
\tau_1(y)  &=&\frac{9}{2\sqrt{2}\pi^2 {f}_1 \hat F}
e^{\Delta/E+\Delta_1/E^\prime}
\int_0^{\omega_{c}}\int_0^{\omega_{1\,c}}
d\sigma d \sigma^\prime~
e^{-\sigma/E-\sigma/E^\prime}
h_1(\sigma,~\sigma^\prime)\theta(\sigma,~\sigma^\prime)
\label{3p1}\\
\tau_2(y)  &=&
\frac{3}{\sqrt{2}\pi^2 \tilde{f}_2 \hat F}
e^{\Delta/E+\Delta_2/E^\prime}
\int_0^{\omega_{c}}\int_0^{\omega_{2\,c}}
d\sigma d \sigma^\prime~
e^{-\sigma/E-\sigma/E^\prime}h_2(\sigma,~\sigma^\prime)
\theta(\sigma,~\sigma^\prime)
\label{3p2}
\eea where
\bea
h_1(\sigma,~\sigma^\prime)&=&\frac{1}{(y^2-1)^{3/2}}\left[\frac{\sigma^2+
\sigma^{\prime 
2}-2y\sigma\sigma^\prime
}{2(y-1)}~+~\frac{\sigma^\prime(\sigma + \sigma^\prime)}{3}\right]
\nn\\
h_2(\sigma,~\sigma^\prime)&=&\frac{1}{(y+1)(y^2-1)^{5/2}}\left[
5\sigma^3-3\sigma^\prime\sigma^2(4y-1)+(2y^2-2y+1)(\sigma^{\prime
3}+3\sigma\sigma^{\prime 2})
\right]~,\nonumber \\
\eea
and
\begin{equation}\chi(\sigma,~\sigma^\prime)~=~\Theta(\sigma^2+\sigma^{\prime\, 
2}-2y\sigma\sigma^\prime)~,\end{equation}
with $\Theta(x)$ the step function.

In eqs.(\ref{3p1}) and (\ref{3p2}) the parameter
$\Delta$ represents the mass difference
between the low lying multiplet $s_\ell=\left(\frac{1}{2}\right)^-$ and
the heavy quark.
The integration region can be expressed in terms of the variables
\bea
\sigma_+&=&\frac{\sigma+\sigma^\prime}{2}\nn\\
\sigma_-&=&\frac{\sigma-\sigma^\prime}{2}\nn
\eea 
and one can choose the  triangular region defined by the bounds:
\bea
0 \leq& \sigma_+&\leq \omega(y)\\
-\sqrt{\frac{y-1}{y+1}} \, \sigma_+&\leq\sigma_-&\leq +
\sqrt{\frac{y-1}{y+1}}\, 
\sigma_+~. 
\eea 
As to the upper limit
in the integration interval for $\omega_+$ we adopt
\begin{equation}\omega(y)~=~\frac{\omega_{1\,
c}+\omega_c}{2\left(1+ \sqrt{\frac{y-1}{y+1}}\right)}\end{equation}
for the two cases studied in this letter (we use, according to the
two-point sum rule analysis $\omega_{c\, 1}=\omega_{c\, 2}$).
 
\begin{figure}
\begin{center}
\epsfig{file=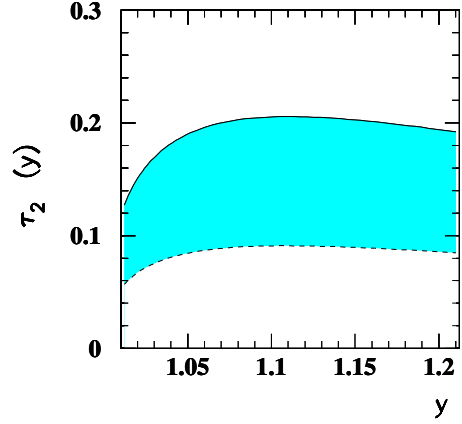,height=9cm}
\label{f:tau2}
\caption{Universal form factor $\tau_2(y)$}
\end{center}
\end{figure}

We use the value $\hat F=0.21$ GeV$^{3/2}$, which is obtained by QCD sum
rules \cite{paver,neubertrev} with $\alpha_s=0$ (the same order which we consider
in the present analysis).
Moreover, we use  $\Delta=0.5$ GeV, with 
the threshold in the $B$ channel  $\omega_c=0.7$ GeV. As for the 
charm channel, we use $\omega_{1\,c}=\omega_{2\, c}=1.6-1.8$ GeV.
 
We can now  numerically determine the form factors $\tau_i$, 
using the above equations. 
The result for the universal function  $\tau_1(y)$, 
obtained within the uncertainties discussed above, is that this function,
in the whole kinematical region relevant for the decays
(\ref{eq:4.1})-(\ref{eq:4.2}), is  less 
than $10^{-4}$, which implies that, in the infinite heavy quark mass limit,
the  semileptonic $B$ transitions into the
$s_\ell= {3\over 2}^-$ doublet have a very small decay width. 
The situation is different for the  universal function  $\tau_2(y)$,
which is depicted in fig.\ref{f:tau2} where the shaded region corresponds
to the results obtained by varying the parameters $\Delta$,  $\Delta_2$,
$\omega_c$ and $\omega_{2c}$ in the ranges quoted above. 
The form factor 
$\tau_2$, at the zero recoil point $y=1$, is in the range 
$\tau_2(1)=0.10-0.20$, with a mild $y-$dependence that can be neglected, 
within the accuracy of the sum-rule method.  Although it is difficult to 
reliably assess the theoretical accuracy of this result, it is interesting 
to observe that a form factor in the range quoted above implies that the 
semileptonic channel is experimentally accessible. 

\section{Semileptonic decay rates}

Using the parameterization of the $B$ matrix elements in 
eqs.(\ref{eq:4.7}) and (\ref{eq:4.8}) we can work out the
expressions of the  widths of the decay modes  
(\ref{eq:4.3}) and (\ref{eq:4.4}), which are respectively given by:
\be
{d \Gamma\over dy} (B \rightarrow D^{*\prime}_2 \ell \nu_{\ell}) =
{G_F^2 V_{cb}^2 m_B^2 m_{ D^{*\prime}_2}^3 \over 720 \pi^3}
(\tau_2(y))^2 (y-1)^{5 \over 2}  (y+1)^{7 \over 2} 
[(1+r^2)(7 y-3) -2 r (4 y^2 -3 y +3)] \label{width2}
\ee
\be
{d \Gamma \over dy} (B \rightarrow D_3 \ell \nu_{\ell}) =
{G_F^2 V_{cb}^2 m_B^2 m_{D_3}^3 \over 720 \pi^3}
(\tau_2(y))^2 (y-1)^{5 \over 2}  (y+1)^{7 \over 2} 
[(1+r^2)(11 + 3 y) -2 r (11 y +3)] \label{width3}
\ee
with $r={m_{D_i}\over m_B}$.
Using $m_{D_3}=2.78$ GeV, $m_{D^{*\prime}_2}=2.74$ GeV 
and $\tau_2(y)=0.15$, we get
\be
\Gamma (B \rightarrow D^{*\prime}_2 \ell \nu_{\ell}) \simeq  
\Gamma (B \rightarrow D_3 \ell \nu_{\ell}) \simeq  4 \times 10^{-18} 
\;\;\;{\rm GeV} 
\ee
and
\be
B (B \rightarrow D^{*\prime}_2 \ell \nu_{\ell}) \simeq  
B (B \rightarrow D_3 \ell \nu_{\ell}) \simeq  1 \times 10^{-5} \;\;\,\;.  
\ee
Therefore, although small, semileptonic $B$ decays to the ${5\over 2}^-$
doublet are within the reach of the running $B$ factories, 
and could be experimentally
observed, since the final mesons, as discussed in the next Section, are 
expected to be rather narrow. 

As for $B$ decays to the  ${3\over 2}^-$ doublet, due to the small value of
the universal function $\tau_1$, the semileptonic widths turn out
to be negligible at the leading order in the $1\over m_Q$ expansion
(a discussion of the role of next-to-leading corrections for semileptonic
$B$ decays to excited charm mesons can be found in \cite{ligeti}).

\section{Remarks on strong decays of orbitally excited charm states}

One might  expect that the states in the multiplets  
$s_{\ell}^P={\frac{3}{2}}^-$ and
${\frac{5}{2}}^-$, being significantly higher in mass than the low-lying
$s_{\ell}^P={\frac{1}{2}}^-$
multiplet, are rather broad. However this should be  only true for the 
$s_{\ell}^P={\frac{3}{2}}^-$
 states. As a matter of fact,  the $J^P=2^-$ and $J^P=1^-$ states 
can decay
into the $0^-$ or $1^-$ heavy meson plus one pion by
$P-$wave transitions, which  implies a kinematical suppression of the order of
$\displaystyle \frac{|\vec p_\pi|^3}{\Lambda_\chi^3}$, where
$\Lambda_\chi\simeq 1$ GeV is the typical chiral symmetry
breaking scale. Taking into account that, for the  charmed mesons, $|\vec
p_\pi|\simeq 0.68$ GeV, we
expect a kinematical phase space suppression, for this 
decay channel, of $\approx 0.3$. 

On the other hand, for the mesons belonging
to the multiplet
$s_{\ell}^P={\frac{5}{2}}^-$, the decay into the low lying heavy meson and
one pion occurs by $F-$wave transitions: the kinematical suppression is  
$\displaystyle\approx \frac{|\vec p_\pi|^7}{\Lambda_\chi^7}$, which
numerically means a reducing factor $\approx 0.07$. Since the decay mode
with one pion in the final state is expected to dominate the decay width,
one may guess that the $3^-$ and $2^-$ mesons 
belonging to the 
$s_{\ell}^P={\frac{5}{2}}^-$ doublet are rather narrow
\footnote{Similar conclusions are reached in \cite{melikhov}.}.

To render these conclusions more quantitative, let us consider the
effective lagrangian describing, in the chiral effective theory for heavy
mesons \cite{falkluke}, the strong couplings of the multiplet $H^{\mu\nu}$
to the pion and the multiplet $H$:
\begin{equation}
{\mathcal L} = \frac{1}{\Lambda_\chi^2}  Tr \left\{
\bar H H^{\mu\nu} \left[ k_1 \{ D_\mu,D_\nu \} {\mathcal A}_\lambda + 
k_2 \left( D_\mu D_\lambda {\mathcal A}_\nu+D_\nu D_\lambda {\mathcal
A}_\mu \right) \right] \gamma^\lambda\gamma_5\right\} + h.c.
\end{equation} 
where $H$ is the 
$s_{\ell}^P={\frac{1}{2}}^-$ multiplet containing the $0^-$ and $1^-$
low-lying states, and $k_{1,2}$ effective couplings;
moreover 
\be
{\mathcal A}_\lambda = 
{1 \over 2} [\xi^\dagger \partial_\lambda \xi - 
\xi \partial_\lambda \xi^\dagger ]
\ee
and $\xi= \exp [i { \vec \pi \cdot \vec \tau \over f_\pi}]$.
Putting $\tilde k=k_1+k_2$, one obtains for the  two-body decay widths:
\begin{eqnarray}
\Gamma(D_3\to D\pi)&=&
\frac{6\tilde k^2|\vec p_\pi|^7}{35\pi f_\pi^2
\Lambda_\chi^4}\frac{m_D}{m_{D_3}}\\
\Gamma(D_3\to D^*\pi)&=&\frac{8\tilde k^2|\vec p_\pi|^7}{35\pi
f_\pi^2\Lambda_\chi^4}\frac{m_{D^*}}{m_{D_3}}\\
\Gamma(D_2^{*\prime}\to D\pi)&=&0\\
\Gamma(D_2^{*\prime}\to D^*\pi)&=&\frac{2\tilde k^2|\vec p_\pi|^7}{5\pi
f_\pi^2\Lambda_\chi^4}\frac{m_{D^*}}{m_{D_2^{*\prime}}}
\end{eqnarray}
with $f_\pi\simeq 132$ MeV. The value of $\tilde k$
is unknown; however, on the basis of QCD sum rule results for similar
couplings \cite{gatto}, one may assume  $\tilde k\in[0.25,0.5]$.
In correspondence to the lower bound in this range we get
\begin{eqnarray}
\Gamma(D_3\to (D,D^*)\pi)&\simeq& 32 \;\;\; {\rm MeV}
\label{w1}
\\
\Gamma(D_2^{*\prime}\to D^*\pi)&\simeq& 15 \;\;\;{\rm MeV}\label{w2}
\end{eqnarray} 
where we have assumed  the mass splitting of $40$ MeV between
the $3^-$ and the $2^-$ mesons in the multiplet.
There are other decay channels contributing to the full widths, but the
corresponding
partial widths are expected to be much smaller: for the decay modes with
one
pion and an excited positive parity $D$ resonance in the final state,
occuring by $D-$wave transitions, 
we estimate a width of 1-2 MeV; for the decay modes
with two pions and a heavy meson in the final state we expect, in the
infinite heavy quark mass limit, a negligible contribution.

We can therefore conclude that  reasonable estimates for the full widths
of the $s_\ell={5\over 2}^-$  resonances are as follows: 
\begin{eqnarray}
\Gamma(D_3)&=&35-140 ~{\rm MeV}\\
\Gamma(D^{*\prime}_2)&=&17-70 ~{\rm MeV} \;\;\;\; ,
\end{eqnarray}
a consideration which suggests the presence of a not too broad peak
in the $D \pi$ and $D^* \pi$ channel in the region of $2.8$ GeV.

This conclusion, together with the result of a branching fraction
of semileptonic $B$ decays to the ${5 \over 2}^-$ doublet of the order
of $10^{-5}$, encourages the experimental investigation 
at the currently running $B$-factories 
as well as at the hadronic facilities.
 
\vspace*{2cm}
\noindent {\bf Acknowledgments}

\noindent
We thank Prof. R. Gatto and Prof. N. Paver for discussions and
collaboration at an early stage of this work. 
(FDF) also acknowledges D\'epartement de Physique Th\'eorique, 
Universit\'e de Gen\`eve, Switzerland, for hospitality, and 
``Fondazione Angelo Della Riccia'' for financial support.

\end{document}